\documentclass[9.3pt,journal,twoside]{IEEEtran}

\usepackage{amssymb}
\usepackage{subfig}

\usepackage{hyperref}

\usepackage{textcomp}
\usepackage{arydshln} 
\usepackage{color} 

\usepackage{diagbox}    
\usepackage{multirow}   
\usepackage{url} 
\usepackage{verbatim}	
\usepackage{longtable}
\usepackage[section]{placeins}

\ifCLASSINFOpdf
\usepackage[pdftex]{graphicx}
\DeclareGraphicsExtensions{.pdf,.jpeg,.png}
\else
\usepackage[dvips]{graphicx}
\DeclareGraphicsExtensions{.eps}
\fi

\ifCLASSINFOpdf

\else

\fi

\begin{document}

	\title{Internet of Behaviors: A Survey} 

\author{Jiayi Sun, Wensheng Gan*, Han-Chieh Chao, Philip S. Yu,~\IEEEmembership{Fellow,~IEEE} and Weiping Ding

\thanks{This research was supported in part by the National Natural Science Foundation of China (Nos. 62002136 and 62272196), Natural Science Foundation of Guangdong Province (No. 2022A1515011861), Guangzhou Basic and Applied Basic Research Foundation (No. 202102020277), and the Young Scholar Program of Pazhou Lab (No. PZL2021KF0023). (Corresponding author: Wensheng Gan)}

\thanks{Jiayi Sun is with the College of Cyber Security, Jinan University, Guangzhou 510632, China. (E-mail: jiayisun01@gmail.com)} 

\thanks{Wensheng Gan is with the College of Cyber Security, Jinan University, Guangzhou 510632, China; and also with Pazhou Lab, Guangzhou 510330, China. (E-mail: wsgan001@gmail.com)}

\thanks{Han-Chieh Chao is with the Department of Electrical Engineering, National Dong Hwa University, Hualien, Taiwan, R.O.C (E-mail: hcchao@gmail.com)} 	
	
\thanks{Philip S. Yu is with the Department of Computer Science, University of Illinois at Chicago, IL, USA. (E-mail: psyu@uic.edu)}	

\thanks{Weiping Ding is with the School of Information Science and Technology, Nantong University, Nantong 226019, China (E-mail: dwp9988@163.com)} 	
	
}


\maketitle

\begin{abstract}	
	The Internet of Behavior is a research theme that aims to analyze human behavior data on the Internet from the perspective of behavioral psychology, obtain insights about human behavior, and better understand the intention behind the behavior. In this way, the Internet of Behavior can predict human behavioral trends in the future and even change human behavior, which can provide more convenience for human life. With the increasing prosperity of the Internet of Things, more and more behavior-related data is collected on the Internet by connected devices such as sensors. People and behavior are connected through the extension of the Internet of Things -- the Internet of Behavior. At present, the Internet of Behavior has gradually been applied to our lives, but it is still in its early stages, and many opportunities and challenges are emerging. This article provides an in-depth overview of the fundamental aspects of the Internet of Behavior: (1) We introduce the development process and research status of the Internet of Behavior from the perspective of the Internet of Things. (2) We propose the characteristics of the Internet of Behavior and define its development direction in terms of three aspects: real-time, autonomy, and reliability. (3) We provide a comprehensive summary of the current applications of the Internet of Behavior, including specific discussions in five scenarios that give an overview of the application status of the Internet of Behavior. (4) We discuss the challenges of the Internet of Behavior's development and its future directions, which hopefully will bring some progress to the Internet of Behavior. To the best of our knowledge, this is the first survey paper on the Internet of Behavior. We hope that this in-depth review can provide some useful directions for more productive research in related fields.
\end{abstract}


\IEEEpeerreviewmaketitle

\section{Introduction}

With the rapid advancement of computer technology, network technology, and sensing technology, the Internet of Things (IoT) \cite{atzori2010internet} has also come into being. The IoT is the ``Internet of all things connected". It is an extension and expansion of the Internet. It combines various information-sensing devices and networks to form a huge network, realizing the interconnection of people, machines, and things at any time and anywhere. Although the term ``Internet of things" wasn't coined until much later, the first device on the web appeared in 1982. Programmers at Carnegie Mellon connected Coca-Cola vending machines to the Internet, so people could check for cold drinks before making a purchase. It is widely regarded as one of the first IoT devices. The IoT got its name in 1999. Kevin Ashton, head of MIT's Automatic ID Lab, first coined the term ``Internet of Things" in a speech to illustrate the potential of RFID tracking technology. In the decades since, devices connected to the IoT have evolved from mobile phones and computers to daily devices like refrigerators and vending machines, and now smart home devices like smoke alarms, speaker boxes, and tablet computers are all connected to the Internet.

The IoT is promoting the transformation of human society from ``information" to ``intelligence", and promoting tremendous changes in information technology and industry \cite{atzori2010internet}. The IoT has brought human society into a highly intelligent era in which things are connected. According to IDC, the global IoT market revenue will reach 1.1 trillion dollars by 2025, with an average annual compound growth rate of 11.4\%, of which China's market share will rise to 25.9\%, making its IoT market the largest in the world. The technical idea of the IoT is to ``connect everything according to demand". All objects are connected to the Internet through information sensing devices for information exchange in order to achieve intelligent identification and management of objects. Since the advent of the Internet, data has been crucial to determining who uses it and which sites people visit. The advent of the IoT provides more data to collect and analyze. The more data one has, the more information one can get about the user's behavior. The use status of IoT devices and data collection can provide valuable insights about relevant user behavior, interests, and preferences, known as the Internet of Behavior (IoB). The IoB attempts to understand the data collected from users' online activities from a behavioral psychology perspective. It seeks to address how data can be understood from a human psychology perspective and how this understanding can be applied to influence or change human behavior.

\textbf{Previous work.} Gote Nyman \cite{nyman2012ib}, a retired psychology professor at the University of Helsinki, was the first to announce the concept of the IoB. He has developed the concept that behaviors can be mined from rich data. Nyman believes it is possible to understand what is about to happen in the connected world by understanding human intentions. The IoB is technically easy to implement but psychologically complex. Statistical studies describe daily habits and behaviors but do not adequately reveal the meaning and context of an individual's life. In 2012, he proposed in \cite{nyman2012pbi} that if each selected and meaningful behavior pattern is assigned to a specific IoB address, it is possible to benefit from the knowledge gained by analyzing these patterns in business, health, and many other fields. Behavior is a psychological characteristic that determines whether a person is willing to cooperate or not. Despite the other characteristics: cognition, emotion, personality, and inter-communication, behavior is responsible for the tendency to act and is highly dependent on the other four criteria \cite{elayan2021internet}. Therefore, we can understand how to influence and treat the person by focusing on their behavior. On October 22, 2019, Gartner made the point that the IoB links people and actions \cite{gartner2019top}. In this way, the IoB will be used to encourage or discourage a particular set of behaviors. Gartner predicts that by 2023, individual activities will be digitally tracked by an ``Internet of Behavior" to influence benefit and service eligibility for 40\% of people worldwide. On October 20, 2020, Gartner released nine key strategic technology trends that organizations need to dig deep into in 2021 \cite{gartner2020top}. Among the nine important strategic technology trends, the IoB is the first one. Gartner proposes the IoB as an extension of the IoT, which focuses on capturing, processing, and analyzing the ``digital dust" in people's daily lives. The term "digital dust" is a good metaphor for the traces that users leave behind in their Internet activities. As the collection of digital dust from everyday life increases, including data that spans both the digital and physical worlds, information can in turn influence users' behaviors. The more data users create, the more digital dust they leave behind, much of which consists of unstructured or semi-structured data. At the same time, with the development of methods for processing and understanding data, we can extract more detailed analyses from the data. They estimate that by the end of 2025, more than half of the world's population will have been exposed to at least one IoB program, whether commercial or government-sponsored.

The IoB has gradually come into our lives. A lot of the data collected by sensors in the IoT is related to specific human behaviors. The IoB attempts to understand this data through behavioral psychology and data analysis, and then applies the relevant findings to influence human behavior. For example, during the COVID-19 pandemic, China established a complete health code program that could record the user's itinerary, contact history, body temperature, and other indicators. Then, the staff can judge the user's health status according to the specific QR code color and impose certain restrictions on specific people to affect their behavior. The smartwatch is one of the most common wearable devices, recording the user's heart rate and sleep status and providing notifications or advice to the user through the program to improve their physical condition. The IoB gives unlimited possibilities to all aspects of life and will have very broad applications in people's future lives. Its potential cannot be underestimated.

\textbf{Current deficiency.} However, the IoB is still in its infancy. The relevant definitions, concepts, and specific applications are not clear, and a lot of content related to the IoB has not been paid attention, which hinders the practical application and development of the IoB. 

\begin{itemize}	
	\item The definition of the IoB is not comprehensive enough. As we know, the IoB emphasizes the collection of data related to people's behavior on the Internet and then predicts or influences people's behavior after classifying and analyzing this data. However, many concepts closely related to the IoB have not been summarized at the current stage due to the complexity and diversity of data sources.
	
	\item Lack of in-depth analysis of specific applications. Unlike other well-established fields, the IoB is still in its infancy. Mature scenarios have been able to put forward perfect modeling schemes and implementation methods for specific applications of the same concept in different scenarios. However, the IoB-based applications have not established a complete application system, and many things are still in the conceptual stage. In the application and development of the IoB, there is still a lot of work to be done, from theory to application.	
\end{itemize}

\textbf{Contributions}: For the above issues, this article proposes the definition and characteristics of the IoB, summarizes the current application scenarios and challenges of the IoB, and puts forward future directions. The contributions of this article are as follows.

\begin{itemize}
	\item To the best of our knowledge, this is the first survey paper on the Internet of Behavior. It first introduces an in-depth understanding of the IoB, that is one of the emerging technologies in recent years.
	
	\item We introduce the concepts related to the IoT and the definitions of the IoB extended by it. The IoB plays an important role in the ternary space of human social space, physical space, and cyberspace (Section 2). 
	
	\item We propose the characteristics (including real-time, autonomy, and reliability) of the IoB in its current stage (Section 3). We also discuss the specific applications of the IoB in various fields, such as business, medicine, education, transportation, and so on (Section 4). 
	
	\item We highlight the future development trend of the IoB. The close integration of the IoB and other cutting-edge technologies will inject new power into the Internet (Section 5). 
\end{itemize}

\begin{figure}[h]
	\centering
	\includegraphics[scale = 0.48]{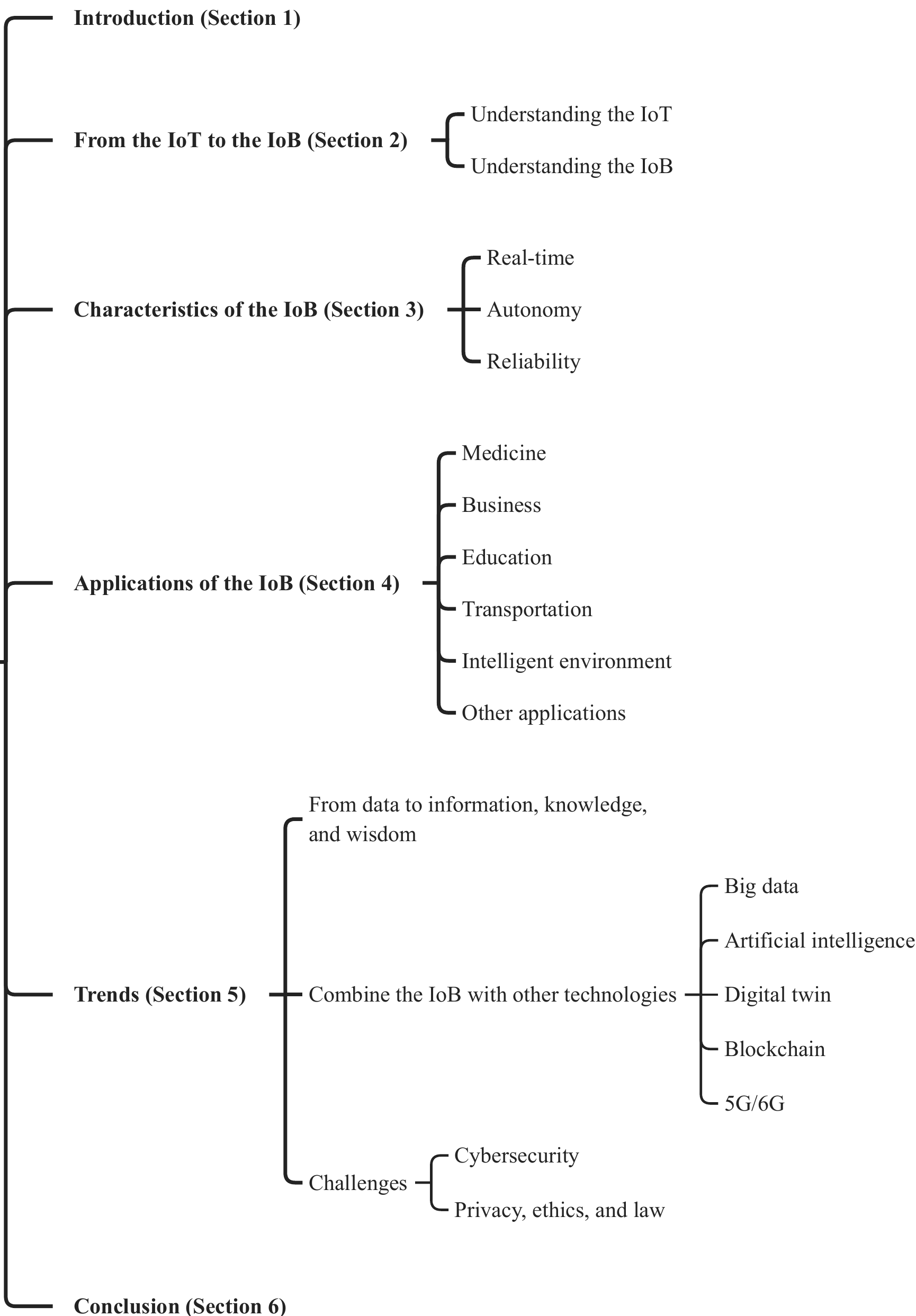}
	\caption{Outline of this article}
	\label{fig:outline}
\end{figure}

The organization of this article is shown in Fig. \ref{fig:outline}.

\section{From the IoT to the IoB}
\label{sec:From the IoT to the IoB}

\subsection{Understanding the IoT}

Originated in the field of media, the Internet of Things (IoT) is the third revolution in the information technology industry. The IoT has evolved from the traditional Internet and telecommunications networks. All physical objects that can be addressed independently are interconnected through the IoT. As the name implies, the IoT is the Internet where things are connected. Through information sensing devices, it connects any object to the network according to an agreed protocol. Objects exchange information and communicate with each other through the IoT. A vast, information-rich network is formed by the combination of the IoT and the Internet.

\begin{figure*}[h]
	\centering
	\includegraphics[scale = 0.46]{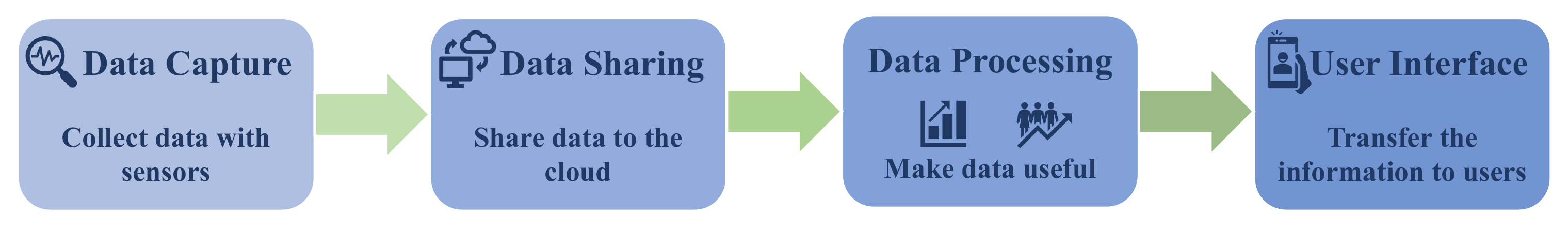}
	\caption{The workflow of the IoT}
	\label{fig:workflowIoT}
\end{figure*}

\textbf{Concept.} In a narrow sense, the Internet of things refers to the network that connects objects to realize the intelligent identification and management of objects. In a broad sense, the IoT can be regarded as the integration of information space and physical space. Everything is digitized and networked for effective information exchange. Information interaction covers a wide range, including between objects, between objects and people, and between people and the real environment. The IoT promotes the comprehensive application of information technology in human society by integrating various information technologies into social behaviors through new service modes. Its purpose is to connect any object with the network according to the protocol and realize an intelligent object. Thus, functions such as identification, tracking, and remote use can be integrated into the object. The complexity of the IoT is expanding and evolving, both in the way devices are connected and in the types of computing, those devices can handle autonomously. This connects devices to each other and storing data in the cloud. The workflow of the IoT mainly includes the following four processes, as shown in Fig. \ref{fig:workflowIoT}:

\begin{itemize}	
	\item Data capture. Using sensors, IoT devices collect data from their environment.
	
	\item Data sharing. Data is sent to the cloud through a network connection, and IoT devices can access the data according to instructions.
	
	\item Data processing. It processes data in the cloud, uses data analytics tools, artificial intelligence, and machine learning to gain actionable and productive insights, and makes the data useful.
	
	\item User interface. It transfers the processed data to the user.
\end{itemize}

\textbf{Process.} The concept of the IoT has been put forward for more than 20 years, since it was first proposed by Professor Kevin Ashton of the Massachusetts Institute of Technology in 1999. In 2000, LG introduced the world's first internet-connected refrigerator, with a screen and tracker to help users track what's in the fridge. However, it failed to win over consumers because of its high cost and the vagueness of the problems to be solved. In 2004, the term ``Internet of Things" began appearing in book titles and circulating in the media. RFID has also been applied in the Savi Program by the US Department of Defense and in business by Walmart. From 2006 to 2008, the IoT gradually gained recognition in the European Union. The first International Internet of Things Conference was held in Zurich, Switzerland in 2008. It was the year that the number of IoT devices surpassed the number of people on earth for the first time. According to the Cisco Internet Business Solutions Group (IBSG), the IoT was born between 2008 and 2009, making more things or objects connected to the network. Google began developing self-driving cars in 2009. In 2013, Google Glass was released, which is a revolutionary advance in wearable technology and the IoT. In recent years, the IoT industry has undergone unprecedented changes due to the improvement of people's living standards and the development of technology. The development of the IoT is becoming easier and more widely accepted. Blockchain and artificial intelligence have begun to be integrated into the IoT platforms. The popularity of smartphones and broadband will continue to promote the further development of the IoT.

\textbf{Future.} The scenarios, applications, and business models of the connection between things and things are diverse. The information in the IoT will be more fragmented. Enterprises based on the data operation platform of the IoT will create more value. The extreme innovation brought by the IoT fully releases people's demand for efficient use of resources. According to Machina Research statistics, the number of global IoT device connections increased rapidly from 2010 to 2018, from 2 billion in 2010 to 9.1 billion in 2018, with a compound growth rate of 20.9\%. In 2025, the number of IoT devices (including cellular and non-cellular) will reach 25.1 billion, and the ``Internet of Everything" will become an important direction for the future development of the global network. According to Gartner, the potential economic value generated by connected data will reach \$11 trillion by 2025, and the global size of the sharing economy is expected to reach \$335 billion, which will make the market even bigger in the era of the interconnection of all things. As an important part of the new generation of information technology, the IoT has expanded from the traditional person-to-person information exchange to the information exchange between things. The means of communication have become more intelligent and convenient.

The in-depth application of the IoT in all walks of life will give birth to numerous new technologies, new products, new applications, and new models. The huge market demand in the future will bring rich opportunities and broad space for the development of the IoT. Companies are using facial recognition, location tracking, and big data to track individual behavior and link it to digital activities such as buying train tickets. The IoT correlates observed operational parameters with expected operational parameters and then instructs the physical entity to execute certain instructions. The Internet of Things (IoT) is now extending to humans through the Internet of Behaviors (IoB).

\subsection{Understanding the Internet of Behavior: extending the IoT}

\textbf{Concept.} The IoB is an extension of the IoT. The IoT, a network of interconnected machines, works with data, information, and the way different devices are connected. To put it simply, the IoT connects all items with the Internet through information-sensing devices to exchange information in order to achieve intelligent identification and management. The IoB also works with the same parameters, but the difference lies in the process of data analysis. It takes user behavior into account and analyzes what specific patterns represent and how they affect the user by trying to understand the behavior and intention. It has the ability to draw conclusions from changing circumstances and make ground-breaking decisions based on those conclusions.

\begin{figure}[h]
	\centering
	\includegraphics[scale = 0.32]{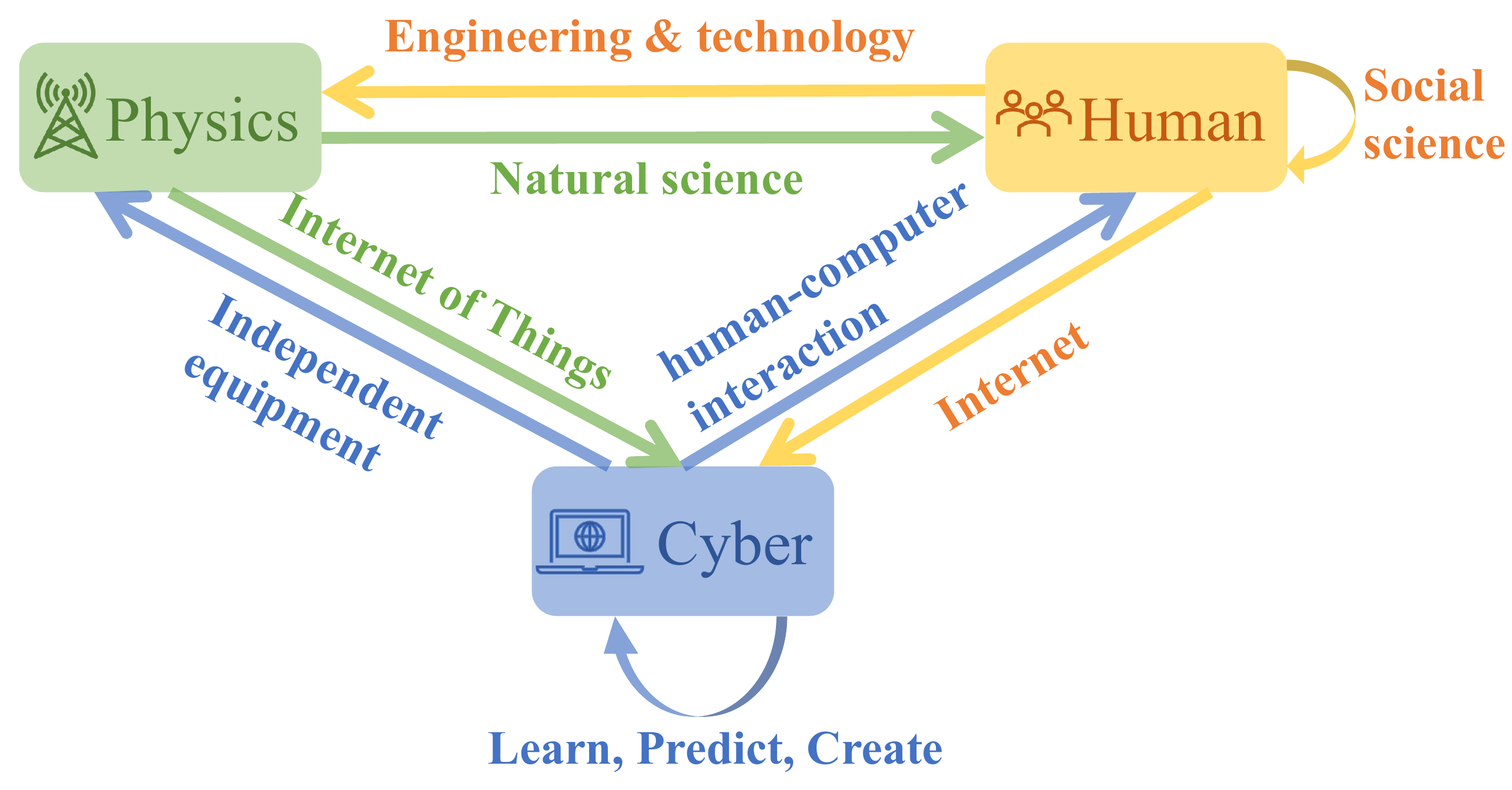}
	\caption{The relationship of H, C, and P}
	\label{fig:relationshipHCP}
\end{figure}

Before the emergence of the Internet, our world was a binary space: Human society space (H) and Physical space (P). With the development of the Internet, the world is forming a new space -- Cyberspace (C). In 2014, it was found that the world is changing from the traditional binary space to the ternary space of human social space, physical space, and cyberspace. Cyberspace is a new dimension that is growing rapidly. In the past 50 years, information has basically been produced by human society. And then, with unlimited connections through the Internet, coupled with information generated in recent years from the physical world like sensors and the IoT, humans have entered the era of big data. In the traditional binary world, there are two types of relationships: one is the relationship between human society and the physical world, and the other is the relationship between human beings and other human beings. After entering ternary society, both the physical world and human society will have a close relationship with cyberspace, and the relationship will become stronger and stronger, as shown in Fig. \ref{fig:relationshipHCP}. The IoB can be seen as a combination of technology, data analysis, and behavioral science. It collects and analyzes data related to human behavior in the ternary society, forming a complete information pathway between the physical world, the digital world, and the human society. In cyberspace, the powerful analytical and predictive capabilities of digital technologies can guide human activities in the real world and promote the common development of ternary space. The IoB is a data analysis network covering ternary space and has become a bridge in the ternary relationship of physical space, human society, and cyberspace.

\begin{figure}[h]
	\centering
	\includegraphics[scale = 0.32]{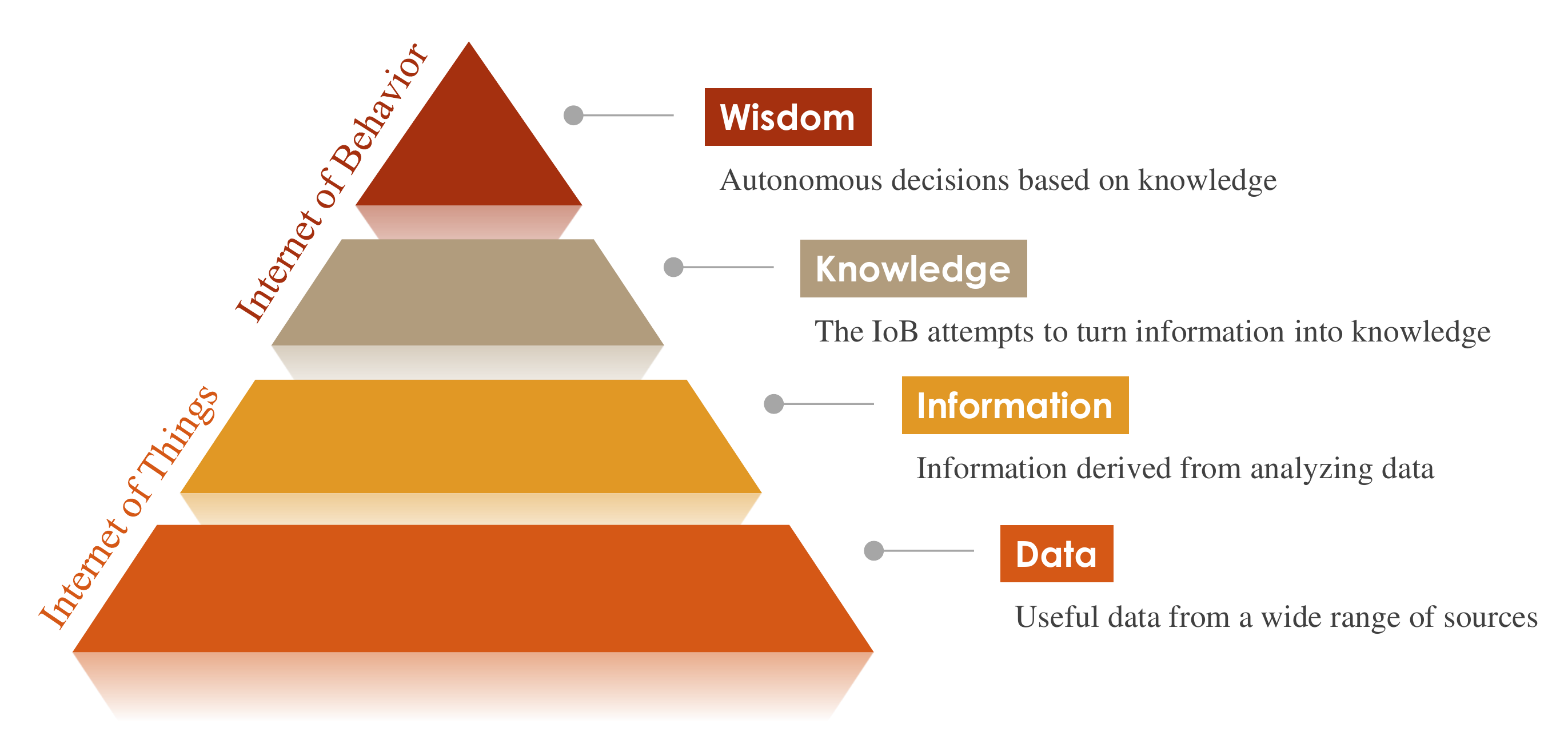}
	\caption{The IoT and the IoB pyramid}
	\label{fig:IoT To IoB}
\end{figure}

\textbf{Development process.} The concept of the IoB first appeared on Gote Nyman's blog in 2012. He proposed that the IoB is similar to the IoT. Unlike the IoT, the behaviors that the IoB deals with do not need to be precise in nature or fixed. They are pre-defined or otherwise specifically declared by someone. These behaviors carry human insights into the addressed behavior pattern. The IoB relies on the wisdom and knowledge of the operator to represent meaningful behavior patterns. Perhaps due to the limitations of technological development, Nyman's novel and cutting-edge views did not receive much attention at that time. Since then, there has been a long-term gap in information about the IoB on the Internet. The interaction between the Internet and Behavior has been concerned constantly, but the related concepts of the IoB are not explicitly mentioned. Until October 2019, Gartner proposed the Internet of Behavior, which links people and actions \cite{gartner2019top}. They emphasize the importance of the IoB in their lives, which makes the IoB gradually come into public view, and the number of articles and discussions related to it on the Internet is gradually increasing. In December 2019, Chrissy Kidd \cite{kidd2019iob} mentioned a pyramid about the IoT and the IoB, as shown in Fig. \ref{fig:IoT To IoB}.

\begin{figure*}[h]
	\centering
	\includegraphics[width=17cm,height=8.5cm,scale = 0.46]{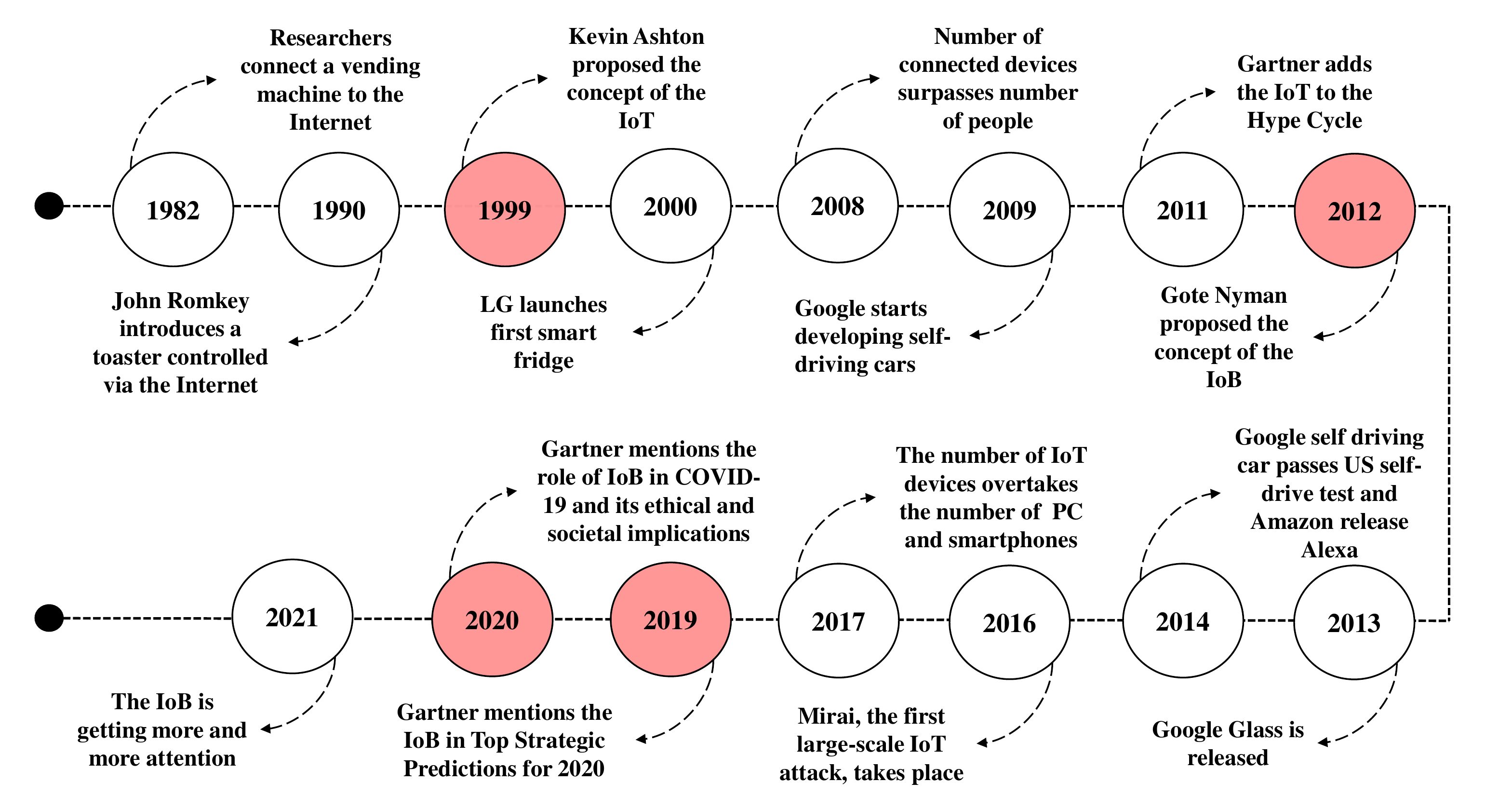}
	\caption{The timeline of the IoB}
	\label{fig:timelineIoB}
\end{figure*}

She regards the IoT as the bottom of the pyramid, and the second layer describes how the network collects data and converts it into information. The third and fourth layers describe how the network uses relevant background knowledge and technology to transform the information into knowledge and wisdom. This pyramid reveals the relationship and differences between the IoT and the IoB. The IoT is the basis of the IoB, a larger network architecture. What the IoB does is take some data from the IoT and attach it to specific human behaviors, while trying to turn the information into knowledge. Then, the wisdom at the top can be obtained by using more complex technologies such as artificial intelligence and machine learning or the background knowledge of behavioral psychology. Nyman mentioned in his April 2020 blog \cite{nyman2020covid} that the IoB can be used for tracking, monitoring, and managing human social activities during the COVID-19 epidemic. It can even predict the possible behavior in a place in order to propose and implement countermeasures as soon as possible. At the same time, the emergence of COVID-19 in 2020 will accelerate the change in the relationship between physical space, human society, and cyberspace. In October 2020, Gartner put forward the forecast again and put the IoB in the first place of the forecast \cite{gartner2020top}. With the gradual increase of available data, the IoB can capture digital dust from all aspects of people's lives, which can be used to influence behavior. As shown in Fig. \ref{fig:timelineIoB}, it describes a brief history from the IoT to the IoB.

\textbf{Workflow.} The IoB is a network of interconnected physical objects that collect and exchange information over the Internet, linking this data to specific human behaviors. As shown in Fig. \ref{fig:workflow of IoB}, its workflow is as follows:

\begin{figure}[!h]
	\centering
	\includegraphics[scale = 0.27]{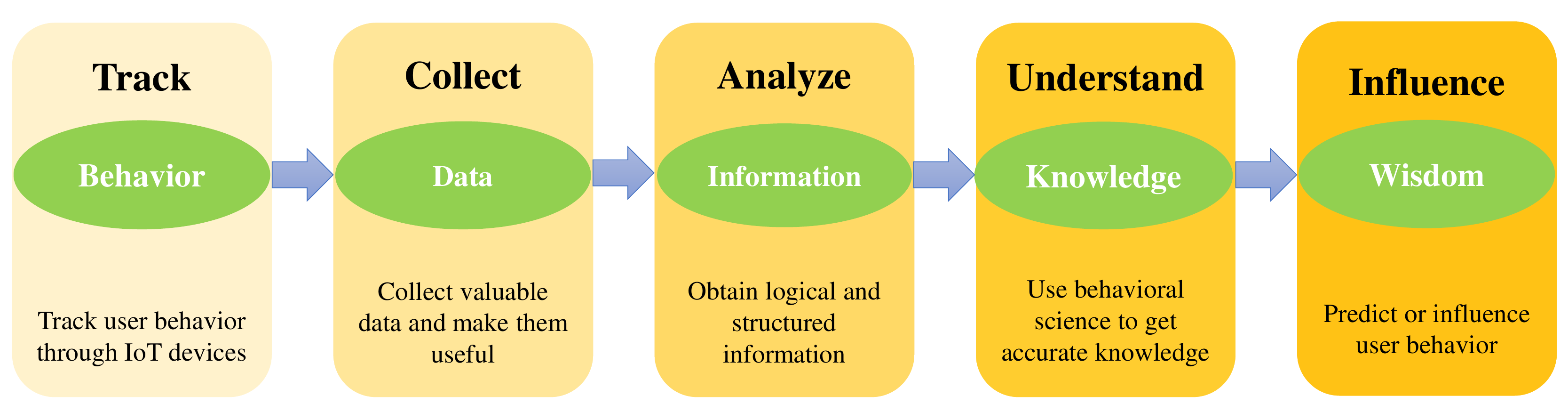}
	\caption{The workflow of the IoB}
	\label{fig:workflow of IoB}
\end{figure}

\begin{itemize}
	\item Use sensors and terminal devices connected to the IoT to track the behavior of users;
	
	\item Collect and make available for analysis all types of useful data generated by IoT devices on an ongoing basis;
	
	\item Use data analysis and machine learning algorithms to sort and analyze the processed data to obtain logical and structured information;
	
	\item Mining useful information with specific patterns and using behavioral science and artificial intelligence algorithms to understand exactly how those patterns affect human behaviors;
	
	\item Use corresponding knowledge to make decisions independently, predict user behavior, or influence user behavior towards the expected direction.
\end{itemize}


\section{The Characteristics of the IoB}
\label{sec:The characteristics of the IoB}

\begin{figure}[h]
	\centering
	\includegraphics[scale = 0.4]{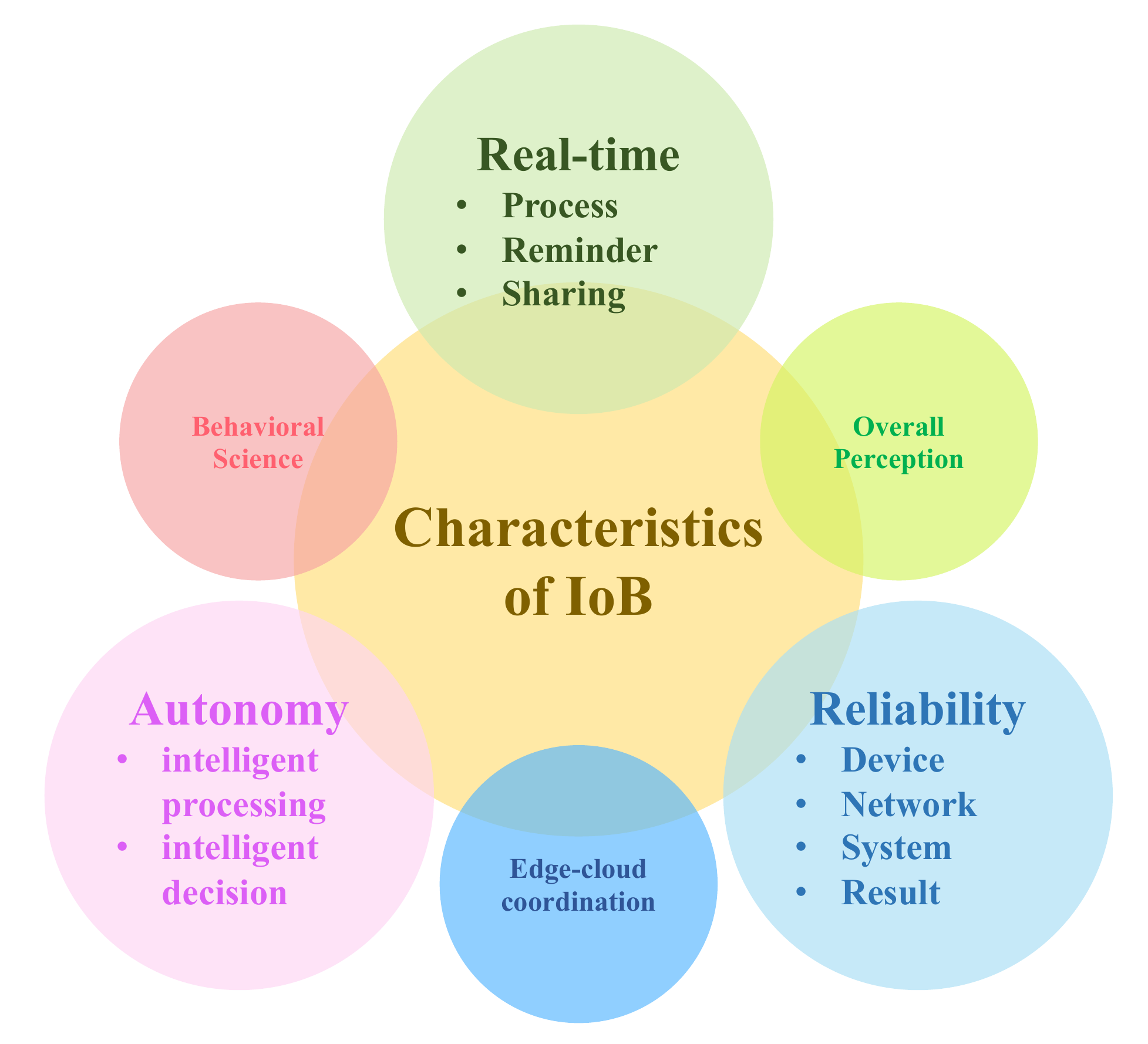}
	\caption{The characteristics of the IoB}
	\label{fig:characteristicsIoB}
\end{figure}

There are some characteristics that the IoB must contain, so that the effectiveness and universality of its application can be guaranteed. Based on similar characteristics, they are divided into several specific categories, as shown in Fig. \ref{fig:characteristicsIoB}. The specific traits are discussed in this chapter.

\subsection{Real-time}

The new generation of information technology is widely used, such as social networks, e-commerce, and the IoB, which are the main sources of big data. In the big data processing of the Internet, the common scenarios are user portraits, recommendation systems, public opinion analysis, and so on. These scenarios do not require real-time performance and can be batch-processed. However, the big data processing platform of the IoB must be a system that receives and processes data in real time, which will allow users to receive real-time feedback about their context. After obtaining the analysis results of relevant application scenario data in due time, users can make wise decisions and reasonable actions based on the feedback \cite{malek2017use}. For the application scenarios of the IoB, real-time analysis, early warning, and decision-making are required based on the collected data. In this case, the applications require high throughput and low end-to-end latency from the system, despite possible fluctuations in the workload \cite{heinze2014cloud}. If there is no real-time data processing, the application scope of the IoB will be largely limited.

The IoB must be able to receive data in real time and ensure that the data is written continuously and reliably. For IoB systems, the data flow is often stable. For example, the generation of sensor data and the writing of sensor data to the database are stable. Therefore, the resources required for data collection can often be estimated. What has changed is querying and analysis, which may consume a lot of system resources and are uncontrollable. The system must allocate sufficient resources to guarantee that data can be written to the system without being lost. Meanwhile, the IoB should process data in real time, requiring real-time streaming computing. When dealing with the infinite data generated in the system, it is necessary to achieve low latency as much as possible while ensuring high accuracy at the same time. Various real-time warnings or predictions are no longer simply based on a certain threshold, but need to be aggregated and calculated by data streams generated by one or more devices in real time. The process is calculated not only based on a point in time but also on a time window. The calculation requirements are also quite complex and vary from scenario to scenario; user-defined functions should be allowed for calculations. Moreover, the processing of real-time data and historical data should be indistinguishable. Real-time data is in the cache. Historical data is stored on a persistent storage medium and may be kept on different storage media depending on the storage time. The system should hide the storage information behind it and present the same interface to users and applications. Whether to access the newly collected data or the old data from many years ago should be the same, except for the different time parameters entered.

As the data is constantly updated, the system must be able to provide real-time reminders and support the data subscription. The same set of data in the system is often used by different applications, so the system should provide a subscription function. As long as there are new data updates, the application should be reminded in real time. The subscription should be personalized, allowing applications to set filter conditions. The real-time sharing of data in the system is also very important. The system should support edge-cloud collaboration, which requires a complete and flexible mechanism to transmit data from edge computing nodes to the cloud. According to the different needs, the mechanism can synchronize the original data, processed data, or qualified data in the cloud in real time. And it must be able to change the strategy or cancel sharing at any time.

\subsection{Autonomy}

The design goal of the IoT is to replace labor with automated equipment, and all equipment at the perception layer, network layer, and application layer can be automatically controlled. In this way, the operating efficiency of the system and the accuracy of the work can be improved, and the maintenance cost can be reduced to a large extent. Similarly, based on the IoT, the IoB can not only achieve automated control, but more importantly, also attain autonomy at the decision-making level. The key to achieving autonomy is the integrated use of technologies in different disciplines, such as the natural sciences and humanities. After simple data processing with data analysis and statistical tools, the relevant algorithms of machine learning and artificial intelligence in the field of computers and background knowledge of behavioral psychology are applied to further analyze and understand behavioral data. It should be able to automatically detect, feedback, and process the relevant data of users and use relevant background knowledge to automatically predict the possible behavior of users in different scenarios.

First, the IoB must be capable of intelligent processing. In order to efficiently process large amounts of data and information, it must be combined with intelligent computing technologies to improve the ability to perceive changes in the physical world and human society. Existing technologies such as cloud computing, artificial intelligence, and data mining are used to process data in the IoB to achieve intelligent decision-making. Second, it also requires the IoB to support flexible multidimensional analysis of data. For data generated by networked devices, various dimensions of statistical analysis are required, such as analysis of the geographic location, model, supplier, and personnel using the device. The analysis of these dimensions cannot be planned in advance and needs to be determined according to specific needs in the actual data and analysis process. Therefore, the big data systems of IoB need a flexible mechanism to add a certain dimension of analysis.

\subsection{Reliability}

The IoB will be connected to production, management, early warning, or other systems in daily life. If any part of the work process fails, it will not be able to provide normal services to end users. For example, the failure of the behavioral early warning system will lead to loopholes in security monitoring, which presents huge security risks. Therefore, the entire operation process of the IoB must be highly reliable, and safeguard measures must be perfected to ensure the reliability of equipment, the network, the system, and the calculation results. Anomaly detection should also be supported, otherwise the service may be interrupted.

\textbf{Device reliability.} For the device, it is necessary to consider its reliability in terms of battery, memory, and computing power \cite{kouicem2018internet}. Most embedded sensors and wearable devices have limited resources in these three areas. The application layer usually does not know the remaining power and service life of the device, and the device is likely to be placed in a dangerous or unreachable place. Thus, while determining the service life of the device as much as possible, it is necessary to realize the non-inductive replacement of the standby equipment. Due to the limited memory and computing power of the sensor, only lightweight encryption methods can be used for the sensor. Consequently, the design of the encryption algorithm and regular firmware upgrades are very important to prevent security holes in outdated firmware. In addition, the sensor has an undetectable problem called ``fail-dirty", which refers to the situation of sending false readings after the sensor fails \cite{jeffery2006declarative}. Raw data should be converted to clean data before the application uses it; otherwise, it will cause invalid work and even serious consequences. As a result, strengthening the monitoring of behavioral Internet devices and troubleshooting the faulty equipment in a timely manner are also necessary measures to ensure device reliability. Mavrogiorgou \textit{et al.} \cite{mavrogiorgou2018capturing} proposed a mechanism to detect the reliability of heterogeneous IoT devices, and users can manage connected devices through a visual interface. Chaudhry \textit{et al.} \cite{chaudhry2020secure} proposed an access control mechanism that not only improves the security of communication between two IoT devices but also improves the efficiency of computing and communication.

\textbf{Network reliability.} A reliable network mainly includes three requirements: low latency, integrity, and security. Reliable transmission refers to the real-time distant transfer of received perceptual information across different telecommunication networks and the Internet in order to actualize information interaction and exchange, as well as carry out various effective processing tasks. In the IoB, the information exchange between the client and the server must be real-time so that the advantages of the IoB can be used to the greatest extent. How to build a reliable network environment is a very important research direction. Low-latency techniques in 5G can greatly increase the data transmission rate, which can achieve perfect data synchronization and smooth user interaction on the Internet \cite{slalmi2021will}. At the same time, network security cannot be ignored, and an efficient and secure network architecture is adopted to ensure the validity of data. Azghiou \textit{et al.} \cite{azghiou2020end} used the reliability block diagram (RBD) paradigm to establish an end-to-end IoT system reliability modeling and analysis framework, providing a new idea for improving the reliability of the IoB network.

\textbf{System reliability.} To achieve a reliable system, high availability must be achieved first. That is, when the system fails, short interruptions are allowed, and rapid recovery is achieved through failure detection and other methods. When improving the availability of IoB systems, business characteristics and requirements should be fully considered, and corresponding failure detection and failure recovery solutions should be automatically deployed \cite{yang2019design}. In terms of disaster recovery, the system must support real-time data backup, remote disaster recovery, software and hardware online upgrades, and online IDC room migration. Most application scenarios of the IoB require quickly accessing the current state or other information of the device for alarm, display, or other purposes. Efficient caching is required to achieve a reliable storage mechanism. The system needs to provide an efficient mechanism so that users can obtain the latest status of all devices or eligible devices. Effective management is necessary for massive amounts of data from different fields and different departments. The system needs to provide flexible data management strategies. A huge system collects a wide variety of data. In addition to the original data collected, there is also a lot of derived data. Each of these datasets has its own characteristics. Some require high collection frequency, some require long retention times, some require multiple copies to ensure higher security, and some require quick access. The system must provide multiple strategies for the specific conditions of different data, and various strategies coexist, allowing users to choose and configure corresponding strategies according to their needs. The final information obtained by the end user relies greatly on the results obtained by the calculation and analysis of the IoB, so the calculation results must be reliable, which ensures the effectiveness of the information and the decision-making based on the information.

\section{Applications of the IoB}
\label{sec: Applications of IoB}

\begin{figure}[!htbp]
	\centering
	\includegraphics[scale = 0.42]{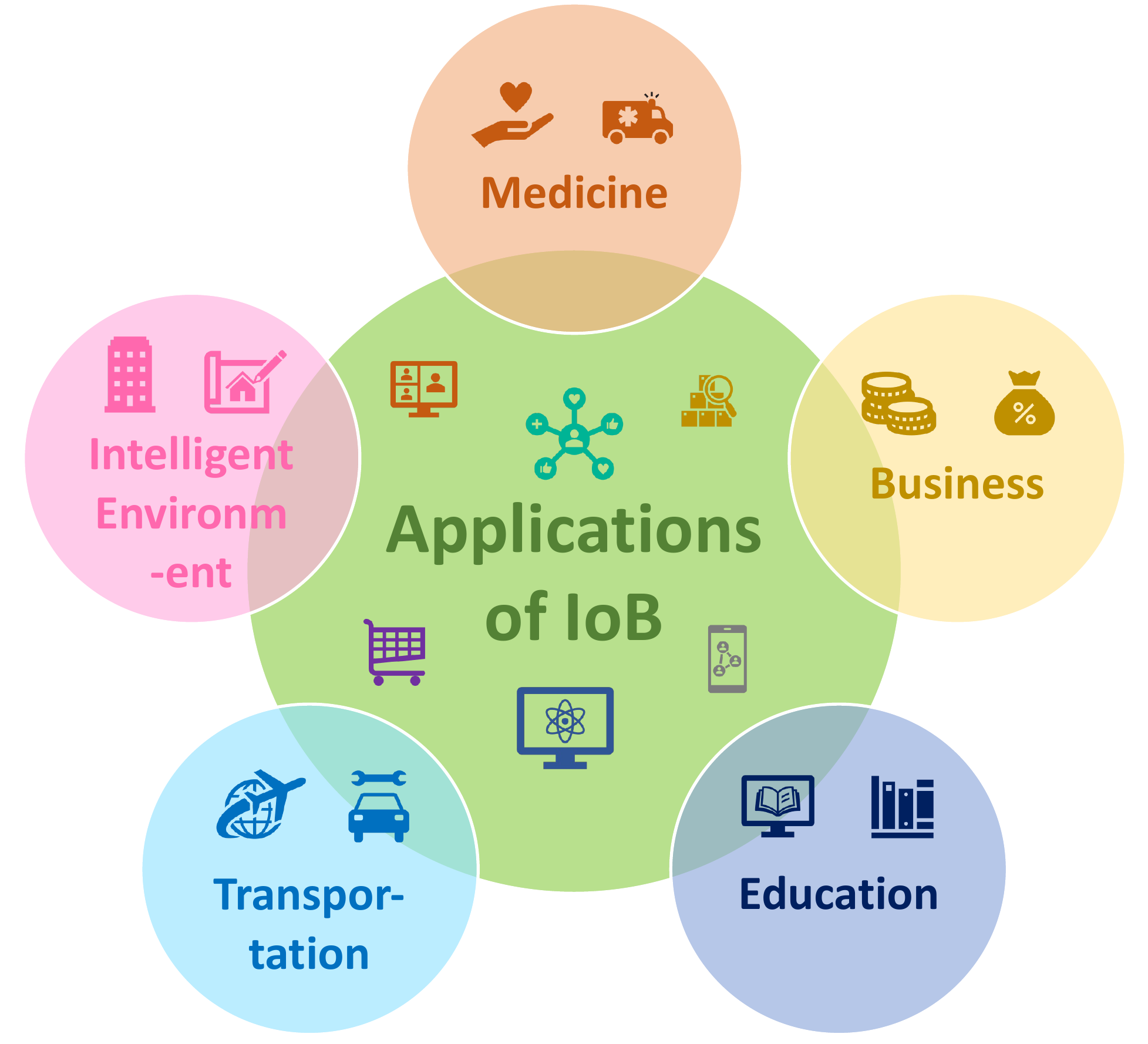}
	\caption{The applications of the IoB}
	\label{fig:applicationIoB}
\end{figure}

This section summarizes the current status of the applications of the IoB in various fields (see Fig. \ref{fig:applicationIoB}).

\subsection{Medicine}

\textbf{COVID-19 and medical services.} Since the outbreak of COVID-19 in 2019, people have gradually adapted to the need to wear masks in public places all the time. Hidden viruses are always a threat to places with high human traffic, such as airports, shopping malls, hospitals, and large companies. Managers have the responsibility to monitor whether everyone wears a mask and keeps a proper social distance from others, but previous management methods have not been effective in detecting people's behaviors. The IoB, a way of collecting and analyzing behavioral data to influence people's behaviors, comes into play at this time. It can conduct real-time monitoring of the behavior of people accommodated in the current place. Using the IoB and computer vision, managers can see whether people have complied with relevant prevention and control regulations and give reminders. Gartner \cite{gartner2020top} pointed out that when the COVID-19 pandemic forced factories to shut down temporarily, employees who returned to the workplace found themselves ``targeted" by sensors or RFID tags. At the same time, these technologies were used to determine whether they washed their hands regularly.

The IoB can be used to track people's movements and contact histories and to trace the source of a new epidemic promptly to prevent and control the spread of the epidemic. In China, people use QR codes to record their visits to public places. Users' social networking and vaccination status will be synchronized with the QR code in real time. The color of the QR code indicates the user's status. Green indicates that everything is normal and that users can go about their normal social activities. Yellow or red indicates that the user is currently at risk of infection and that their social activities will be reasonably limited to prevent the further spread of the virus. This approach has played a significant role in the long-lasting prevention and control of the epidemic. When an epidemic outbreak occurs, the dispatch of medical resources needs to be carefully considered. When conducting nucleic acid tests in high-risk areas, the IoB can monitor the status of medical staff and residents at the same time, analyze the flow of people queuing in real time, and arrange the corresponding number of medical staff according to the specific situation to improve the detection efficiency.

In addition, special care should be given to those who have special circumstances to be tested. When hospitals deal with large-scale emergency epidemic events, the IoB can predict and analyze the number of patients arriving at the hospital within a certain period. By registering patients online, hospitals can prepare ahead of time. The online app will also promptly give feedback to users that the hospital's current free medical resources can be used to receive new patients. If the resources are insufficient, the nearest available hospital will be recommended to patients. In addition, after collecting various health data from users, remote health testing, medical consultation, and digital diagnosis are all possible \cite{mukati2021healthcare}. The spread of the virus can also be controlled under the premise of ensuring timely and effective medical treatment for patients. In the process of treating patients, doctors' operations can be recorded in real time by IoT devices, which can not only accumulate experience for follow-up treatment but also standardize doctors' operations to a certain extent and relieve the tense doctor-patient relationship when necessary.

\textbf{Healthcare.} At present, the most common IoT device used to monitor human health is the smartwatch. The Apple Watch from Apple Inc. can detect human heart rate, sleep duration, blood oxygen level, noise level, and other health indicators. By connecting to the user's mobile phone app, it visually displays the results to the user and gives certain reminders or suggestions. For example, users will be reminded to exercise after sitting for a long time. A fast heart rate will remind you to take a deep breath to relieve tension. Users can open the corresponding mode to record their motion state during exercise and get the next improvement plan after exercise. Intelligent software uses sensors or smartphones to continuously monitor the user's activities, predict and analyze the user's physical state, and issue early warnings to users and doctors in a timely manner when abnormalities are detected \cite{gupta2017novel}. At the same time, specific fitness and diet plans can be made for users based on the analysis results. A mobile app called Clue\footnote{https://helloclue.com}, which tracks a woman's mood, sleep, weight, and other data to predict her menstrual cycle, could make life easier for women.

At present, the proportion of the elderly in the world's population is gradually increasing. The elderly are facing greater health risks because they are more susceptible to age-related physical diseases and cognitive impairments. For example, Mild Cognitive Impairment (MCI), which is prone to occur in the elderly, can easily develop into a more serious condition (such as Alzheimer's disease) if it is not detected and treated when the condition first appears. Therefore, the use of sensors to monitor the behavior of the elderly, combined with background knowledge of clinical medicine, can help doctors find behavior changes caused by diseases as early as possible, and provide patients with customized treatment measures \cite{almeida2017iot}. In a nutshell, the IoB is already making its way into medicine. Behavior detection is mainly carried out in two ways: active and passive. The active way is for people to express their actions and thoughts. The passive way is for automatic collection by personal devices or smart sensors \cite{nyman2020covid}. Obviously, the former has great potential to promote the study of human behavior. It is designed to ensure timeliness and accuracy, and can achieve better results in terms of personal needs than any AI or ML-based system.

\subsection{Business}

\textbf{Marketing.} A lot of research in psychology can give a lot of inspiration to marketing. A very common example is the ``Foot In The Door Effect", which refers to a psychological effect that allows others to accept smaller requirements first to encourage them to accept larger requirements. Applying this kind of human psychology to marketing can get better results. A marketer will put forward a small request that people can or are willing to accept, such as a free experience of his product, and more marketing opportunities will follow, allowing them to achieve their marketing goals step by step. Therefore, it is natural to combine psychology and behavior analysis to understand the useful data collected on the Internet \cite{lo2018blending}. In the past, consumers' attitudes and responses to products or services were determined by the consumers themselves, who filled out various questionnaires or participated in group surveys. This method costs a lot of human and material resources, and the accuracy of the results often fluctuates because of the limited sample space and different respondents. As a combination of technology and behavioral science, the IoB can help companies understand, analyze, and even influence people's behaviors and attitudes toward their products and services \cite{munnia2020big}. The unique advantage of the IoB in the business field lies in its ability to analyze consumer purchase records on any network \cite{javaid2021internet}. Through the use of IoB devices, the aggregation of information and the unique perspective of psychology can bring huge commercial value. Companies mostly utilize the IoT and the IoB to monitor and attempt to modify people's behavior in order to achieve their objectives. Personalization, according to marketers and behavioral scientists, is critical to a service's performance \cite{kidd2019iob}. The goal of the IoB application in business is to correctly understand and apply the collected data to produce and promote products. The company uses the information understood by the IoB to produce products that are popular with users, optimize the user experience, and promote the company's services and products in order to seek greater benefits.

\textbf{Advertising.} YouTube\footnote{https://www.youtube.com}, which we are very familiar, uses behavioral analysis to improve the user's experience. It obtains users' preferences by analyzing users' viewing history and only recommends to them what they are interested in. This method has been widely used on various websites and mobile applications. The operators all try to seize a lot of user time. Such a situation is often encountered in daily life: if the keyword ``birthday" is mentioned in WeChat, when you open the shopping software on your mobile phone, you can see that similar products for birthday gifts are recommended on the home page. If you have searched for some products in the shopping software and then switched to another social software, the position of the advertisement will show the similar product that you just searched for. These mobile phone applications collect users' browsing traces, use algorithms to analyze and predict users' future needs, and combine consumer psychology to influence users' shopping behaviors to gain more benefits \cite{varnali2021online}.

Meanwhile, based on the results of the analysis of this behavioral data, more online advertising can be formulated to improve the accuracy of advertising. The precise advertising targeting technology on the Internet is the result of analyzing user behavior data and is also an application of the IoB. It uses the user's cookie, relying on the huge Internet user behavior database of the search engine, to conduct a personalized analysis of almost all the user's online behavior. According to the analysis results, the technology targets the target audience according to the needs of the advertising provider, conducts one-to-one communication, and provides multichannel delivery  \cite{nigam2016iot,aksu2018advertising}. In the offline store, the face recognition system is combined with the coffee shop. The system recommends suitable drinks to the user by analyzing the user's gender, age, mood, and other data. This system is also suitable for personalized advertising in offline stores.

\subsection{Education}

The effectiveness of education is directly related to students' behavior and habits and teachers' teaching methods. Therefore, it is necessary to deeply analyze all kinds of behavioral data in the education process and make the educational methods more efficient after drawing relevant conclusions and measures.

\textbf{Teacher and student.} The IoB can be used to collect data about students' learning behaviors, such as concentration and performance. By analyzing this data, the system can come up with personalized learning methods and effective suggestions for different students. One of the major benefits of the IoT in education, according to Itai Asseo, strategic innovation executive at Salesforce, is individual and unique contact with students \cite{bagheri2016effect}. On the other hand, with the help of wearable devices, students' health can be monitored regularly. Using their physiological signals and behaviors to judge whether they have a tendency to be depressed or hurt themselves, teachers can prevent such behavior from happening in time. What is more, the system can take a student's medical history, monitor the progress of a student's disease, and send an alert to school staff and their parents if an emergency occurs. In the educational process, it is necessary for teachers to keep abreast of each student's performance and feedback. In smart classrooms, the IoB can collect data on students' performance during the learning process through sensors and face recognition. Through the feedback information of students, teachers can accurately change the plan and method of future courses so that the teaching method is generally suitable for every student in the class \cite{aldowah2017internet}. The combination of more networked devices, such as tablets, and data analysis technology can help teachers monitor students' attendance and activities during tests and classwork and ultimately provide more flexible and personalized teaching.

\textbf{Smart classroom.} First, the smart classroom system can control the usage of power resources and water resources by collecting data on students' behavior. With the help of the traditional IoT ecosystem, combined with the behavior data of people on campus, it can ensure the effective detection and control of the use of energy and water resources, thus creating an environmentally friendly campus environment. Second, the system can make schools safer. Real-time monitoring of people's activity tracks is important for schools, laboratories, and other crowded places, and how to easily monitor students' movements is a problem to be solved. Embedded technology provides a solution to this problem. Currently, NFC technology allows for real-time monitoring of classroom usage status and the display of specific information on a screen outside the classroom. In addition, embedding RFID tags into students' ID cards can easily record students' behavioral data, such as attendance \cite{chang2011smart}.

\subsection{Transportation}

In the field of transportation, using the IoB to analyze drivers' behavioral data can not only timely remind drivers of dangerous driving behaviors, reduce the frequency of road accidents, and ensure the safety of passengers, but also evaluate drivers' safe driving coefficient and customize insurance costs. Furthermore, the results of the analysis are used to train and adjust driver behavior to make it more in line with the values of environmental protection.

\textbf{Safe driving.} Environmental conditions and driver behavior are closely related to traffic safety. A dangerous road environment or improper driver operation will eventually lead to serious traffic accidents, in which the driver's behavior plays a decisive role in traffic safety. Therefore, it is very important to use the IoB to predict unsafe driving patterns. Three driving behavior monitoring and prediction systems based on the IoT are summarized in \cite{abbas2020driver}, which are multi-sensors, configuration-based, and cloud-based architectures. Such systems can predict fatigue and alert drivers when their physical or emotional signals are abnormal. Multidimensional assessments, such as the deployment of sensors and smartphones, can detect drivers' dangerous driving behaviors in a timely manner so that they can quickly correct their poor driving. The systems can also motivate good driving and discover hidden behavioral trends.

\textbf{Eco-driving.} In fact, green driving and safe driving may overlap to a large extent \cite{young2011safe}, and they are consistent in terms of goals. Safe driving requires drivers to avoid dangerous driving operations and reduce the risk of a collision. Eco-driving recommends avoiding speeding and aggressive driving. Vehicle equipment can continuously monitor driving conditions, and the monitored parameters usually include fuel consumption, speed, and road and traffic conditions \cite{huang2018eco}. By analyzing the monitoring data, the system can provide the driver with driving performance feedback and improvement suggestions, and help the driver get rid of bad driving habits, thereby reducing fuel consumption and carbon dioxide emissions. Massoud \textit{et al.} \cite{massoud2019eco} proposed a human driving analysis algorithm. It can train drivers' driving behaviors in a digital environment and help drivers improve fuel efficiency by using serious games.

\textbf{Usage based insurance.} By recording the driver's behavior data, the information asymmetry between the insured and the insurance company can be reduced, so that detailed risk differentiation can be made according to the driver's true risk level. The traditional method is to use statistical data to determine specific driver risk factors. This method does not achieve the purpose of personalized insurance premiums. Using personalized data collected by IoT sensors to determine insurance premiums not only gives drivers better control over their premiums and encourages safe driving, but also cultivates customer loyalty for suppliers through regular contact. Arumugam \textit{et al.} \cite{arumugam2019survey} proposed a solution for calculating personalized premiums based on driving behavior. The solution is to identify abnormal driving behavior by analyzing the driver's behavioral and emotional factors while driving. If any abnormal driving behavior is recognized, a real-time voice alert will be sent to the user via a smartphone. The insurance company will use the collected data to calculate personalized premiums.

\subsection{Intelligent environments}

\textbf{Smart building.} Energy management is a very important direction in the development of intelligent buildings. What is more, one of the main factors affecting energy consumption in buildings is the behavior of occupants \cite{yohanis2012domestic,santin2009effect,hong2017ten}. Therefore, architects all over the world are interested in the behavior of occupants. For decades, social scientists have been studying the behavior of occupants, especially in the fields of user behavior, attitudes, and personal or family consumption patterns \cite{sovacool2014we}. Researchers generally believe that occupant behavior is the main source of uncertainty in the process of energy consumption prediction \cite{hoes2009user}. Compared with changing people's behavior, carrying out technological reforms while maintaining users' original habits has more research significance. Based on a deep understanding and accurate prediction of human behavior, researchers can improve building technology and energy technology, thereby improving the energy efficiency of buildings. For example, building simulation can optimize the overall performance of the building by modeling user behavior \cite{hoes2009user,yan2015occupant}. Using IoT devices in the office building to obtain data related to user behavior can identify high-energy consumers and low-efficiency behaviors \cite{rafsanjani2020towards}. By intervening in these incorrect behaviors, energy efficiency can be improved. Elayan \textit{et al.} \cite{elayan2021internet} proposed a power-saving system based on the IoB and explainable artificial intelligence, and the experiment showed that the system successfully realized power-saving by influencing users' consumption behavior.

\textbf{Smart home.} The IoB can bring great convenience to people's lives. This can be well reflected in smart homes, where automated systems provide comfort to occupants \cite{alaa2017review}. The parameters in the traditional IoT home system are pre-defined by developers or users and do not have the feature of automatic adaptation. When the IoT is expanded to the IoB, insights about human behavior are integrated into smart homes. Smart devices analyze people's activity data to obtain knowledge about people's frequent behavior patterns and living habits, and then adjust themselves to better adapt to those lives. For example, the function of a smart refrigerator is no longer just to help users keep food fresh but also to learn the type, structure, and inventory of the food in the refrigerator through long-term observation. Once the stock is insufficient, the smart refrigerator will prompt what to add and even place an order automatically according to the user's habits.

\onecolumn
\begin{longtable}{|c|p{3cm}|p{14cm}|}
	\caption{The applications of the IoB}
	\label{AppofIoB}
	\\
	\hline
	\textbf{No.} & \multicolumn{1}{c|}{\textbf{Applications}} & \multicolumn{1}{c|}{\textbf{Description}} \\
	\hline
	\endhead
	\hline
	\endfoot
	
	1 & Track behavior and habits & The IoB connects the data collected on the Internet with human activities, and obtains knowledge about human behavior and habits from it. This knowledge plays an important role in the development of various industries, and almost all industries can redesign the supply chain based on the useful knowledge \cite{javaid2021internet}. \\ \hline
	
	2 & Forecast trends & Based on the results of analyzing human behavior data, combined with data mining and machine learning-related algorithms, the trend of behavior can be reasonably predicted. The prediction can be used to develop early warning systems (e.g., health, social security, etc.), adjust services related to human flow, design business, management strategies, and other fields. It has a high application value and broad application prospects. \\ \hline
	
	3 & Training & The IoB comprehensively records people's learning behaviors, such as data during training or learning, to provide learners with suggestions and methods that are more conducive to their learning according to the analysis results of this data. For example, a blockchain-based online language learning system can monitor students' daily learning and automatically evaluate their behavior. This system can not only alleviate the tasks of teachers but also make credible and reliable assessments of students' learning behaviors \cite{sun2021blockchain}. In sports training, through mobile applications and tracking wearable devices, it helps golfers to correct existing hitting techniques and learn new techniques to improve their playing skills \cite{Ksenija2021iob}. \\ \hline
	
	4 & Save energy & The use of energy is often related to human behavior, so the behavioral internet plays an important role in saving energy. Fuel efficiency can be improved by improving the driving behavior of drivers \cite{huang2018eco,massoud2019eco}. Occupant behavior is one of the main factors affecting building energy consumption. Through analyzing and modeling occupant data, energy efficiency and occupant comfort can be improved \cite{hong2016advances}. In terms of daily life, by counting people's activity patterns in public places (e.g., airports, campuses, companies, etc.), and rationally planning the use of power supply and water supply facilities, resources can be effectively saved. \\  \hline
	
	5 & Personalized service & In many fields, such as healthcare, advertising, and education, the importance of personalization is self-evident. After fully analyzing and understanding users' behavior and habits, the IoB can mine frequent patterns related to human behavior. Using this information, personalized recommendations and services can be formulated, which not only improves user satisfaction and facilitates users' lives but also improves the work efficiency of service providers and reduces work costs. \\ \hline
	
	6 & Generate strategy & The IoB can help managers or operators generate reasonable strategies. During the COVID-19 pandemic, the factory used sensors or RFID tags to supervise workers washing their hands regularly, and computer vision was used to remind workers to wear masks \cite{gartner2020top}. For marketing companies, understanding customer habits and preferences is extremely important. The company can innovate and optimize its products according to the direction that suits the users in order to seek greater profits. \\ \hline
	
	7 & Eliminate information asymmetry & Collecting users' behavioral data in daily life can reduce information asymmetry between the parties to the transaction. For example, the sensors in the car are used to record the driving behavior and habits of drivers. In this way, different drivers are evaluated and graded, and car insurance at different prices is also provided \cite{arumugam2019survey}. Similar to auto insurance, the healthcare industry's usage-based insurance can also be realized by using sensor equipment \cite{samuel2015internet}. Insurance companies can use wearable devices to track fitness activities to reduce premiums. It can also be used to monitor users' purchases. Too many unhealthy foods may increase premiums. \\ \hline
	
	8 & Change the culture & The IoB is not allowed to track all behavioral data, especially data related to personal privacy. Unreasonable use or abuse of behavioral data can easily cause personal privacy infringements and leave users with potential safety hazards. At the same time, it is not conducive to the further development of IoB. Therefore, the culture in the IoB era and a reasonable legal system need to be established in time \cite{kidd2019iob}. \\	 \hline
	
	9 & Credit score & Internet giants collect social data, browsing data, online shopping data, entertainment data, and other useful information from the huge number of users on the Internet. Through the information, commercial institutions will analyze user behavior and psychological characteristics, calculate the user's credit score, and provide certain life conveniences according to the credit score. This will guide the user's next behavior and increase their dependence on their products. \\    \hline
	
\end{longtable}
\twocolumn

Meanwhile, the smart home system can provide support for daily health care. Using health monitoring technology, it is possible to collect and analyze the occupant's relevant physiological conditions and daily activity data, as well as give personalized recommendations. Users can check their health signals and information on a real-time and archival basis when needed \cite{li2013smart}, to improve the effectiveness of care and treatment. Physically challenged individuals and the elderly usually need extra care in their daily lives. The smart home system can solve the problem of no one taking care of them \cite{demiris2008technologies,robles2010applications,sokullu2020iot}. The design and implementation of related functions of the system should follow the needs of users \cite{demiris2008technologies}. In addition to directly setting relevant parameters by users, the daily monitoring and analysis of their behavior are of great significance. Monitor the user's life behavior through inconspicuous sensors and provide relatives and caregivers with necessary data \cite{sokullu2020iot} to ensure that a remote emergency alarm can be issued when a dangerous situation occurs, so that a timely rescue can be carried out. In addition, based on the implementation of smart home systems, telemedicine has also become a reality \cite{li2013smart}. Through appropriate means, doctors can provide necessary health-related services for more people.

The other application directions of the IoB are summarized here, as shown in Table \ref{AppofIoB}. Note that the common disadvantages and challenges of the Internet of Behavior in these mentioned specific applications in real-life will be discussed in detail in Section \ref{sec:Trends}.

\section{Trends}  \label{sec:Trends}

With the rapid development of data and information technologies such as 5G, cloud computing, artificial intelligence, the Internet of Things, and big data, global data traffic is increasing exponentially. Cisco has predicted that there will be 29.3 billion connected devices globally by 2023, up from 18.4 billion in 2018, and about half of those connections will support various IoT applications \cite{cisco2020cisco}. The IoT vendors also generate and capture vast amounts of data about people's interactions with these devices. After the formation of digital society, everything in life can be described based on data. These standardized descriptions can provide a basis for accurate judgment and intelligent decision-making. Although human behavior data on the Internet constitutes only a small portion of the massive data, its significance cannot be overstated. Human beings are the carriers of all social activities, so it is of great research significance to understand people's intentions toward behavior, predict the development trend of behavior, and change behavior in the expected direction. The outbreak of the COVID-19 epidemic has further promoted the digital development of society. From healthcare to social life, many social activities have been affected by the epidemic and are accelerating their expansion towards digitalization. This change not only brings huge opportunities for development but also brings many challenges and potential risks. The first is the technical level. The IoB, while combining knowledge of psychology and sociology to understand human behavior, also needs to use big data analysis, artificial intelligence, blockchain, and other technologies to process massive amounts of data efficiently and accurately to obtain meaningful results. In the specific implementation of the IoB, there may be many difficulties that need to be broken through. There is still a lot of work to be done before truly understanding human behavior, from current ``information" to obtaining effective ``wisdom". The second is the moral and legal aspect. When disseminating and using a large amount of data related to behavior, it is necessary to think about how to protect users' privacy and security. It requires us to change the existing cultural system based on experience and establish new systems and rules applicable to the IoB era. Only in this way we can provide users with secure services and ensure the long-term development of the IoB.

\subsection{From data to information, knowledge, and wisdom}

The most fundamental and valuable part of the IoB is behavioral data. These datasets are the beginning of any analysis. Only by ensuring reliable data sources, accurate processing and analysis can be carried out, and effective results can be obtained. This part mainly introduces the capabilities that the IoB needs to have from the three main processes for converting data into useful information.

\textbf{Collection and management of multi-source data.} The sources of behavioral data are different. Compared with data analysis from a single source, the data processed by the IoB realizes comprehensive data fusion in which the data is multi-port, multi-industry, and multi-source. From sensors to mobile terminal equipment and application software, user behavior data has great differences in the data source, data structure, generation time, use place, code protocol, etc. Therefore, the collection of behavioral data must be comprehensive, accurate, and real-time. The IoB should ensure the integrity and reliability of data sources. And the system needs to conduct effective and intelligent management of different platforms, different structures, and different types of data, and build databases suitable for different types of data. Through the accurate and real-time collection of raw data and the standardized management of data with different structures, effective data analysis is realized.

\textbf{Standardized processing and prospective analysis.} First, the standardized processing of behavioral data is the basis of data analysis. For the raw data with different structures, the IoB needs to design a reasonable data standardization plan and quantify different behavior patterns in a clear and scalable way. This improves the efficiency of the data analysis process as well as the accuracy and usability of the analysis results. Second, the IoB should also be forward-looking. Data mining algorithms enable data analysis to better understand the data. Through data cleaning and processing technology, combining psychology and sociology, reasonable modeling is carried out to judge the predictability of data mining results.

\textbf{Visualization and shareability of results.} First and foremost, the visualization of analysis results is important. Combined with the digital twin, the visualization of data analysis results can expand the application scenarios of IoB. Secondly, the wide range of data sources and the standardization of data processing determine the shareability of analysis results. The IoB can break the dilemma of information islands, fully stimulate the potential value of behavioral data, and promote unified and interconnected data and resource sharing among industries and departments. Authoritative, legal, multi-source data resources, and reliable data analysis processes ensure that the displayed content and published data are authoritative.

\subsection{Combine the IoB with other technologies}

\begin{figure}[h]
	\centering
	\includegraphics[scale = 0.6]{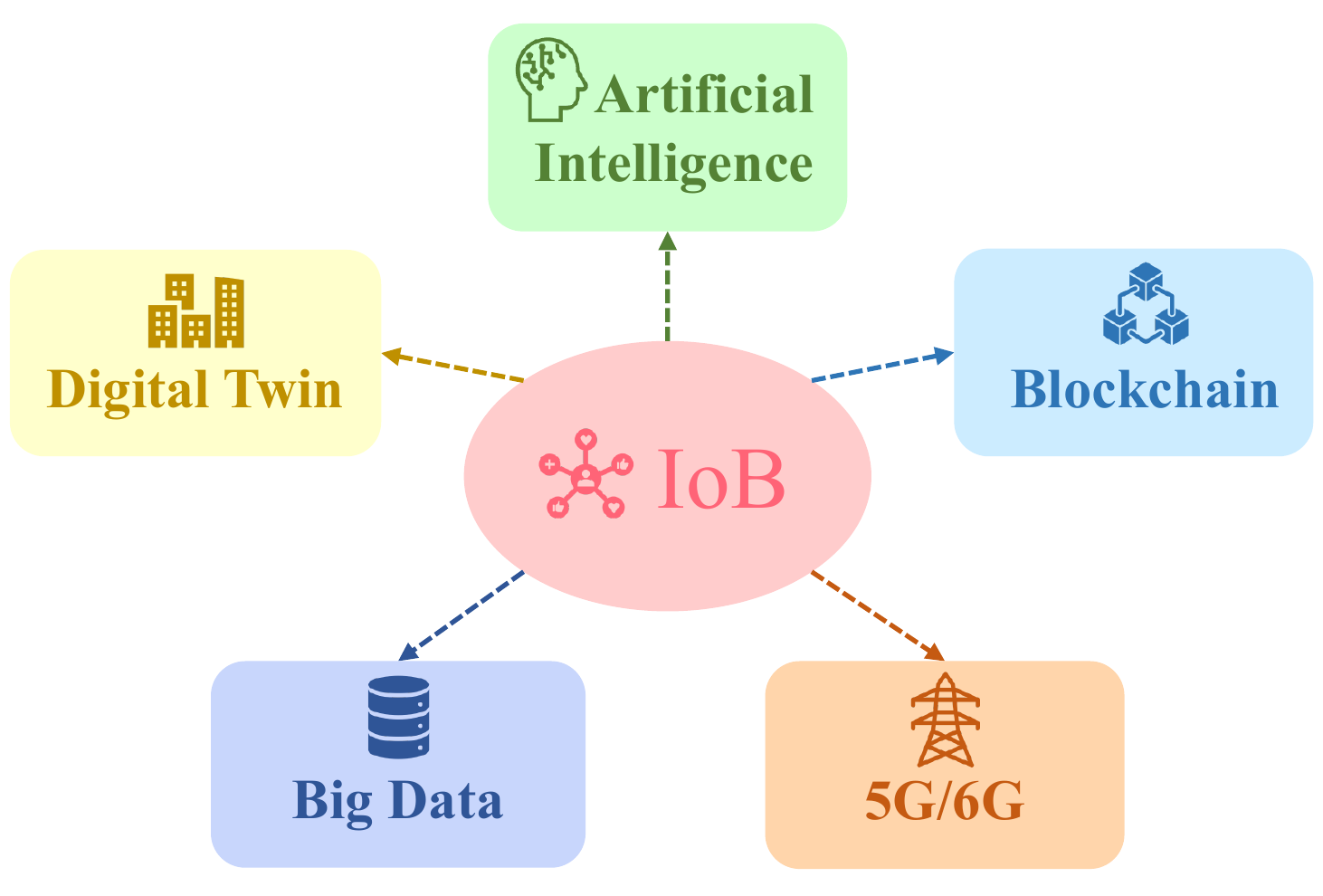}
	\caption{The combination of advanced technologies and the IoB}
	\label{fig:combination}
\end{figure}

The combination of advanced technologies (e.g., big data \cite{gan2019survey,sun2022big}, Metaverse \cite{sun2022metaverse,chen2022metaverse,lin2022metaverse}, artificial intelligence \cite{xu2019explainable}, and blockchain \cite{chen2021construction}) and the IoB can not only overcome the framework and system defects of the IoB, but also inject new vitality into its development (see Fig. \ref{fig:combination}).

\textbf{Big data.} In the era of big data \cite{tsai2015big,fournier2022pattern}, the IoB will generate and capture a large amount of valuable and useful data. In our opinion, big data and computing power are the foundation of the IoB. As an important technical means of decision support and process control, data mining will play a key role in promoting system intelligence and providing more convenient services \cite{de2016iot,shadroo2018systematic}. Towards data-driven intelligence, data mining can successfully extract interesting and potentially useful patterns or rules from data \cite{gan2017data,gan2021survey,gan2020huopm}. It can build an effective prediction or description model for data and provide useful information for the subsequent decision-making or evaluation process. Data mining will run through the entire IoB data process to achieve decision-making, control, and prediction \cite{tsai2013data}. With their automated and intelligent decision-making capabilities, they also can provide security solutions such as identity verification or malware monitoring for the IoB. Therefore, data mining in the IoB should be an indispensable link.

\textbf{Artificial intelligence.} AI aims to create an intelligent machine that can react in a way similar to human intelligence. It can perform various functions, such as perception, learning, reasoning, and solving problems autonomously. Meanwhile, it can find predictive insights into the future based on a large amount of historical and real-time data. First, for the analysis of a large amount of behavioral data in the IoB, AI plays an important role as a powerful analysis tool, that can provide scalable and accurate data analysis in real-time. Since it can analyze past and real-time data at the same time, AI can notice abnormal situations and make reasonable responses. Moreover, as a science that incorporates psychology, AI will have an advantage in processing data related to people. Explainable artificial intelligence (XAI) \cite{xu2019explainable,gade2019explainable} can describe its purpose, principles, and decision-making process so that every device or person in the IoB can understand its execution process. This allows users to better understand what the application is doing, thereby alleviating trust issues to a certain extent. It can provide convenience in self-driving cars, smart homes, and smart medical care. Secondly, the IoB also faces various challenges, such as security, privacy, intelligent decision-making, and so on. AI can provide new and reliable solutions to these problems. Machine learning, deep learning, and neural networks are often used to automatically solve complex problems. In addition, Human In The Loop (HITL) \cite{nunes2015survey} in machine learning can inject new impetus into the development of the IoB. HITL integrates personal judgment into the AI process. Human responses not only determine the final answer, but will also be used to improve the machine learning model, allowing the machine to become smarter, so that it can solve problems that only humans can understand or solve at first.

\textbf{Blockchain.} With the rapid growth of the number of devices connected to the network and the amount of useful information, the risk of private data leakage will gradually increase. Therefore, the security and privacy issues of the IoB cannot be ignored. Reasonable technologies should be introduced to address potential flaws in the IoB and ensure the reliability of the service and the privacy, integrity, and confidentiality of the behavioral data. The decentralization, autonomy, auditability, data encryption, and other functions of the blockchain \cite{zheng2018blockchain,li2020survey} can help solve most of the structural defects in the IoB. The use of blockchain can realize the transformation from the original centralized architecture to a distributed architecture, which can effectively solve the single point of failure problem and at the same time improve the scalability and fault tolerance of the system \cite{dai2019blockchain}. The use of smart contracts on the blockchain in the IoB can not only satisfy the need for secure communication between devices but also realize the autonomous operation of smart devices, eliminating centralized authorization or human control \cite{reyna2018blockchain}. Using the data immutability of the blockchain, sensor data can be tracked and reviewed at any time, which improves the reliability of the data. In short, blockchain is expected to play an important role in the management and protection of devices and data in the IoB.

\textbf{Digital twin.} Physical entities such as sensors in the IoB are one of the main sources of behavioral data. Only if the reliability of the raw data is ensured, the subsequent processing and analysis will be effective. So it is very important to monitor the state of physical entities. As an emerging technology in the Metaverse \cite{sun2022metaverse,chen2022metaverse}, the digital twin \cite{tao2018digital,liu2021review} digitally and accurately models the IoB system, synchronizes state information in real time, and establishes a connection mapping between sensing data and the digital model, enabling the digital model to reflect the status of the IoB system in the real world in real time. In this way, anomaly detection, prediction, and early warning can be carried out for the system. While monitoring and identifying potential problems with physical entities, it can also predict the future of the entity through analysis. The IoB collects data through networked devices, while the digital twin is used to construct and manage data for analysis and optimization using AI and machine learning algorithms to achieve better decision-making and a higher level of automation \cite{kaur2020convergence}. The combination of the digital twin and the IoB can reduce the complexity of the system, make physical entities more intelligent, and make the interaction between users and the environment smoother.

\textbf{5G/6G.} The IoB needs new performance standards in the future. It should be able to support massive connections, achieve wireless communication coverage and ultra-low latency, and ensure safety and reliability. The application of 5G to the IoB not only improves transmission efficiency and shortens decision-making time, but also provides powerful real-time analysis capabilities \cite{shafique2020internet,pham2020survey}. The current 5G communication framework has achieved high throughput, low latency, and high quality. These capabilities effectively guarantee the service quality and network reliability of the IoB and enable the real-time transmission of massive data. With the combination of 5G, edge computing, and artificial intelligence, smart devices in the IoB can realize autonomous interaction and sharing. Many applications on the market now have higher requirements for network communication architecture \cite{chettri2019comprehensive}. These applications are data-intensive, computationally-intensive, and delay-sensitive. Therefore, wireless communications and networks need to carry out technological innovations for 6G to achieve higher data transmission rates, lower latency, and more precise positioning \cite{dogra2020survey}. The distributed vision of 6G is also a very important point, as it can effectively alleviate the various problems faced by the current centralized management of IoT devices. Furthermore, cloud computing \cite{stergiou2018secure,de2019foundations,sadeeq2021iot} can provide storage and computing power for the IoB. Thus, all these will greatly promote the development of the IoB.

\subsection{Challenges}

\textbf{Cybersecurity issues:} The IoB collects and analyzes a large amount of useful data, and this data often reveals the behavior and habits of users. The digital information is exactly the key knowledge that the attacker wants to obtain. With the expansion of the IoB and the increase of access devices, the risk of important information loss or tampering due to attacks on the network or devices will increase. Therefore, the security of the network and equipment must be considered at the beginning of the construction of the IoB. 

First, it should ensure that the hardware and software of the equipment can be updated in time. Updating the hardware and software of networked devices in a convenient and efficient manner is a problem that needs to be considered. If updates cannot be released regularly or the cost of updates is too high, it will lead to an increase in the risk of attacks and increased difficulties in the deployment of devices. 

Second, the equipment must be thoroughly inspected for safety. When there are large-scale networked devices, it is difficult to monitor all the devices. At the same time, it is difficult to judge whether networked devices have been attacked. How to obtain the status information of the terminal equipment in real time and determine whether it is safe is also a problem that needs to be overcome.

Last but not least, predicting and preventing attacks in time is an important way to solve security problems. Cybercriminals are looking for new intrusion techniques. At present, common network attacks include SPoF \cite{huang2019towards}, DDoS \cite{kolias2017ddos}, MiTM \cite{canteaut2013sieve}, phishing \cite{alabdan2020phishing}, etc. These attacks will continue to exist in the IoB and need to be solved \cite{cao2020survey}. In this case, it is necessary to find vulnerabilities, fix them, and learn to predict and prevent new threat attacks \cite{saied2016detection,vishwakarma2020survey,basit2021comprehensive}. It is very complicated to adjust and apply monitoring and analysis tools in the IoB, that involves the real-time processing of data from various terminal devices.

\textbf{Privacy, ethics, and law:} The most basic and important issue in the IoB is people's behavioral data. This data reveals all aspects of a person's living habits, and the problem of privacy leakage will follow. Hence, several issues need to be considered regarding how to protect private data. The first is how to determine the appropriate extent and scope of data collection. When the IoB enters people's lives in an all-around way, people's behavior will be recorded in various ways, such as software, sensors, locators, etc. The type, extent, and scope of data collection need to be considered. This not only reduces the burden of the whole system but also protects people's privacy. More and more homes are becoming smarter through network connections. Although home intelligence is a good thing, the risk of personal privacy leakage is even greater. Users' IP addresses, residential addresses, contact information, and other private data are linked in a closer way. Even if only one item is exposed, other information may be leaked and accessed by hackers, causing problems with privacy disclosure. One way to ensure data privacy is to remove biometrics when necessary. The second is how to ensure the reliable storage of data. A large amount of data is stored on the server or in the cloud and used for calculations and analysis. If the data is not protected properly, it may be stolen by hackers and sold to other companies that violate data privacy rights. Therefore, the behavior data should be encrypted, and the database should also take preventive measures, such as user identification, failure recovery, and strengthening server security. The third is how to use data reasonably. Currently, many companies share data with other departments or subsidiaries. For example, the acquisition of software by Google, Facebook, and Amazon brings users of one application into the entire ecosystem. This will bring significant legal risks. These issues have almost no legal protection. The insurance company may check the specific information and data in the user's social media on the basis of predicting the user's safe driving coefficient. This is likely to be a violation of personal privacy. The use of specific behavioral data necessitates the creation of relevant laws and regulations in order to protect users' legitimate rights and interests, such as privacy rights. Besides, there are no clear rules for the ethical use of information. The IoB needs to change the cultural and legal norms established for big data. If countermeasures are considered in response to the above problems, the IoB will be more acceptable to people.

\section{Conclusion}  \label{sec:conclusion}

The IoB is a relatively new and rapidly increasing idea that aims to analyze data acquired from consumers' online activities through the view of behavioral psychology. From a human psychology standpoint, the notion aims to address the issue of how to correctly comprehend data and how to utilize that understanding to affect people's behavior. The IoB has become something we are increasingly aware of in our daily lives. It ties the resulting data to relevant behaviors by combining existing technologies that focus directly on individuals, such as facial recognition and location monitoring. The IoB will expand and evolve in tandem with the IoT, since it is an extension of the IoT. Meanwhile, the IoB will evolve into an ecosystem that will define human behavior in the digital world in a few years. This will greatly change our lives. Meanwhile, there are many challenges to be solved. The application of behavioral psychology to big data analysis is still in its initial stages; the induction of concepts and the establishment of models are not mature. At the current stage, the IoB cannot accurately understand the hidden intentions behind human behavior, which leads to poor accuracy in predicting behavioral trends that need to be improved. IoB can be used in broader and more stringent fields once the accuracy and efficiency of behavior prediction have reached a certain level. At the same time, the gradual development of IoB will cause many problems, including technical and non-technical issues. In terms of technology, it is necessary to adapt or design IoB-compliant standards and architectures to promote more stable and reliable operation. For non-technical aspects, relevant privacy protection measures and security laws and regulations need to be improved in advance for the arrival of the IoB era.

\ifCLASSOPTIONcaptionsoff
  \newpage
\fi

\bibliographystyle{IEEEtran}
\bibliography{SurveyIoB}

\end{document}